# AERO FIGHTER - 2D GAMING

Z. Ahmed

*Abstract*— **Designing and developing quality based computer game is always a challenging task for developers. In this paper I briefly discuss aero fighting war game based on simple 2D gaming concepts and developed in C & C++ programming languages, using old bitmapping concepts. Going into the details of the game development, I discuss the designed strategies, flow of game and implemented prototype version of game, especially for beginners of game programming.**

## I. INTRODUCTION

Game programming is one of the famous well market branch of software engineering. Since the concepts of programming started initiating, developing and improving their selves, the programming of game applications is also running parallel to the programming abilities [2]. In beginning of visual programming, when there were Assembly, Fortran, Pascal and C like languages since then programmers are developing game applications. In this short paper I am not discussing any new programming concept or anything which currently used and running for latest game programming. The major emphasis of this short paper is the presentation of an old programming concept, especially famous for game programming and 2D graphical user interface designing .i.e., Bitmap.

Bitmap concept is compromises of 16 to 256 color bitmaps, mostly used in used in DOS and windows environments [1]. Bitmapping is consists of binary strings, combinations of 0s and 1s in the form of nibble, byte and words as presented in Table 1. Mostly the concept of Bitmapping for game programming is adopted in C language, the individual file type of bitmap based images is bmp. Moreover even if we look back in our near past then we will find that bitmap based games were very often used in mobiles as well.

| WORD | | |
|---|---|---|
| | byte | |
| | | nibble |
| 1100001 | 1100 | 1001 |
| | | |

Table .1. Binary Table

## II. AERO FIGHTER

### A. Game Introduction

In this short paper as I mentioned earlier, I am discussing game programming but with the implementation of the concept of Bitmap. To give an example of game programming in C programming language using the concept of bitmapping, I am going to discuss one my own developed game in Borland C programming language .i.e., Aero Fighter. Aero Fighter is a two dimensional game, designed and implemented using bitmap and 2D game programming concepts. The whole game consists of following elements (See Figure, 1)

1. Three stages (game plans)
2. User fighter aero plane
3. Enemy fighter aero planes
4. Bullets.
5. Game information
    a. Speed Dial
    b. Height Dial
    c. Plane Destroyed
    d. Time Printing
    e. Scroll Printing
    f. Life Printing
    g. Title Name to the user while playing the game.

The ultimate goal of the user is to play the game and try to kill enemy planes and complete three stages to end the game. The level of complexity and difficulty is increased in each level of the game.

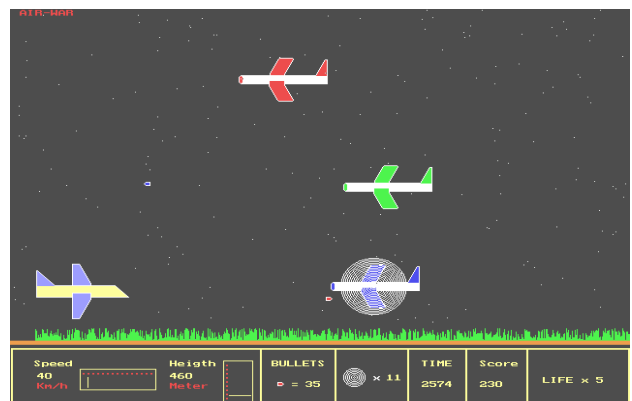

Figure .1. Aero Fighter

### B. Aero Fighter – Game Flow

The flow of the game as shown in Figure 2 consists of following elements .i.e.,

1. Included and declared header files.
2. Declared several six arrays for the making of the images of four different planes (One of them is user plane and other are enemy planes), and two different bullets.
3. Declared Prototypes

Z. Ahmed is presently working as University Assistant with Mechanical Engineering Informatics and Virtual Product Development Division, Vienna University of Technology, Getreidemarkt 9/307 1060 Vienna Austria (phone: 004315880130726 - email: zeeshan.ahmed@tuwien.ac.at and Zeeshan.ahmed@hotmail.de).

4. Declared class 'blast_plane' to call three functions 'void blast', 'void s_sound' and 'sound'.
5. Declared second class 'blast_bullet' to inherit sound function from the 'blast_plnae'.
6. Declared class 'capture_design' to call the functions void 'front', 'capture_image' and 'show_grass' to present and ground grass movement.
7. Declared main method.
    a. In the start of the main method call classes.
    b. Next step get the VGI drivers.
    c. Get pixels on the screen randomly.
    d. Initialize the values 'user cols' 'user rows' and similarly 'enemy1 cols' 'enemy1 rose'. Est.
    e. Start the main while loop of the program with condition continue till press ESC.
    f. Apply three if conditions for three stages of the game, like if user scores 20 then jump to next stage or if user score 40 then jump to next stage.
    g. Similarly apply condition for the end of the game.
    h. Then 'cout' the 'Speed Dial', 'Height Dial', 'Plane Destroyed', 'Time Printing', 'Scroll Printing', 'Life Printing' and 'Title Name'.
    i. Move the pixels on the screen by the help of loop, and call the 'show_grass ()'.
    j. Apply the conditions and assign the ASCII value of arrow keys to move the plane in the screen on the pixels and keep it on the screen.
    k. Put the image of the user plane and user bullet.
    l. Put the images of the enemy planes and enemy bullet.
    m. Apply various condition and loops for the making of the collisions between the user plane with enemy plane and enemy's bullet, user's bullets with the enemy planes and enemy bullet.
    n. Similarly the collisions between the enemy planes and enemy plane's bullet with the user plane and user plane's bullet.
    o. Apply conditions for the movement of enemy planes in the screen.
    p. Now, 'main' method comes to the end and the process of the making void function whose prototypes is declared in the start and used in the main.
    q. Function of showing grass is declared( show_grass()).
    r. Function of the main front screen declared, and in this function we also declare the 'Speed Dial', 'Height Dial', 'Plane Destroyed', 'Time Printing', 'Scroll Printing', 'Life Printing' and 'Title Name'.
    s. Capture the image of the user and enemy planes.
    t. Declare the function of 'plane blast', which produces the circle when ever, a plane or a bullet blasts.
    u. Declare the functions of the sound 'sound1 ()' and 'sound'.

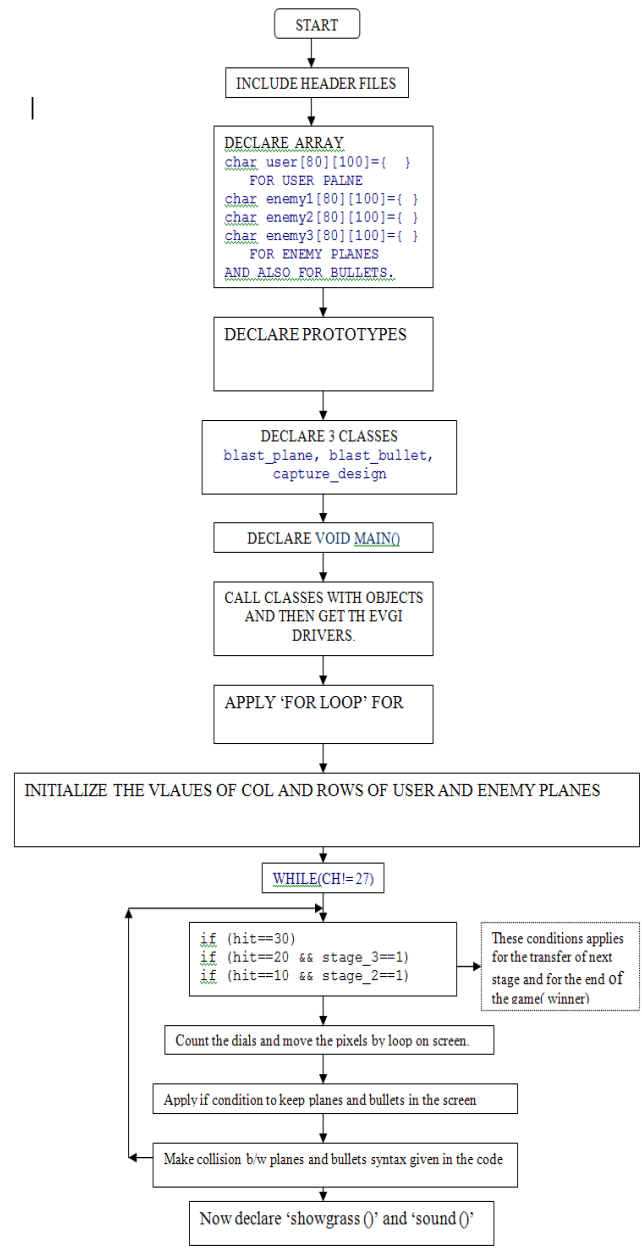

Figure .2. Aero Fighter – Game Flow

III. CONCLUSION

In the end I just want to conclude with the statement that the development of future technologies including new games, tool and methodologies etc. will never end but keep on increasing but it doesn't mean we should compromise on learning the basic and major concepts of development concepts of game programming. This short paper can be an example for new game programming developers.

ACKNOWLEDGEMENT

As the part of acknowledgement I would like to thanks Society of Advancement of Computer Science, Punjab Institute of Computer Science, University of Central Punjab to allow me to participate in their Advance Programming Competition Fall 2001.